\documentstyle[prd,aps,eqsecnum]{revtex}

\begin{document}

\title{\begin{flushright}
{\normalsize KUL-TF-96/10\\ hep-th/9605218\\ To appear in Phys. Rev. D}
\end{flushright}
On the determination of anomalies in supersymmetric theories}
\author{Friedemann Brandt\cite{Friedemann,newadd}
and Jordi Par\'{\i}s\cite{Jordi,newadd}}
\address{Instituut voor Theoretische Fysica\\
         Katholieke Universiteit Leuven\\
         Celestijnenlaan 200 D\\
         B--3001 Leuven (Heverlee), Belgium}
\maketitle
\begin{abstract}
We develop an efficient technique to
compute anomalies in
supersymmetric theories by combining
the so-called nonlocal
regularization method and superspace
techniques. To illustrate the
method we apply it to a four dimensional
toy model with potentially
anomalous $N=1$ supersymmetry and prove
explicitly that in this
model all the candidate supersymmetry
anomalies have vanishing
coefficients at the one-loop level.
\end{abstract}


\setcounter{page}{1}

\section{Introduction}

\hspace{\parindent}%
Supersymmetric quantum field theories
have many remarkable
properties. In particular quantum
corrections are usually better
under control in such theories
than in others
due to nonrenormalization properties
implied by supersymmetry.
However, it is not clear from the outset
whether the supersymmetry of
a classical theory survives as a symmetry
of the quantized
theory, due to the lack of consistent regularization
methods which manifestly preserve supersymmetry
in perturbation theory.
Nevertheless, supersymmetry `miraculously'
appears to be preserved in standard supersymmetric theories.

An indirect but powerful and regularization independent
tool to investigate whether or not supersymmetry can be
anomalous consists in an analysis of the supersymmetric
analog of the Wess--Zumino consistency condition \cite{wz}.
Nontrivial solutions to this consistency condition are candidate
supersymmetry anomalies whereas the absence of such solutions
indicates that  supersymmetry is not anomalous.

The consistency condition for supersymmetry anomalies,
in combination with the
usual Wess--Zumino consistency condition in the case of
supersymmetric gauge theories, has
been studied already
for various $D=4$, $N=1$ globally 
supersymmetric models, see e.g. \cite{past,rigid}, and recently
also for minimal supergravity \cite{sugra}.
It turns out that whether or not candidate supersymmetry anomalies exist
depends decisively on the way supersymmetry is represented on the fields,
i.e. on the structure of the supersymmetry multiplets present in
the model in question. For standard representations, such as multiplets that
can be described in terms of unconstrained or chiral scalar superfields,
one finds that candidate anomalies for supersymmetry itself do not
exist. However, this does not exclude the existence of supersymmetrized 
versions of other candidate anomalies such as ABJ chiral anomalies
in super Yang--Mills theories.
Moreover, there are non-standard representations of supersymmetry
("non-QDS-representations" in the terminology of \cite{rigid}) which
do give rise to candidate anomalies for supersymmetry itself.

When the cohomological analysis alone is not sufficient to exclude
candidate anomalies due to the existence of nontrivial solutions
to the consistency condition (for supersymmetry or other symmetries),
one has to check by an explicit calculation whether or not these
candidate anomalies have vanishing coefficients. To that end
one needs an appropriate regularization method.
One of the main disadvantages of most of the regularization
methods designed for supersymmetric theories
is the lack of a consistent
implementation of the superspace techniques
\cite{bible0,wb}
--one of the main tools in supersymmetry--
at the regularized level
\cite{bible0}. This drawback, somewhat
analogous to the
dimensional regularization troubles
when dealing with chiral theories,
becomes then relevant in analyzing
the presence of anomalies
in the model under consideration.
Indeed, naive manipulations in
superspace may lead to inconsistencies
or ambiguities when computing
divergent expressions, making
impossible to detect and calculate
(unambiguously) such anomalies.
It would thus be desirable to design a
method in which superspace computations were
unambiguously defined.

In this paper we develop a new efficient technique to
investigate anomaly issues in supersymmetric theories.
It combines naturally
superspace techniques, which
facilitate the perturbative calculations in supersymmetric
theories considerably, with the so-called
nonlocal regularization \cite{emkw,j95},
which has already been successfully
used to compute one \cite{hand92} and
higher loop anomalies \cite{j95} in
other (nonsupersymmetric) theories.
Among others, the method allows to
check whether or not supersymmetry
itself is anomalous.
We illustrate this by applying the method
to a four dimensional supersymmetric
toy model whose supersymmetry is
potentially anomalous, as cohomological
results indicate \cite{rigid}.

The paper is organized as follows. First
we describe our method in section \ref{reg}.
To that end we briefly recall
the basic concepts of nonlocal regularization,
emphasizing its use to determine anomalies,
and describe how superspace techniques are
naturally implemented in it.
In section \ref{model} we introduce the
toy model and present
its candidate supersymmetry anomalies.
In section \ref{comp} we then apply our
method to this toy model
and prove the absence of supersymmetry
anomalies at the one-loop level.
Three appendices finally collect our conventions.

\section{Nonlocal Regularization of Supersymmetric Theories}
\label{reg}

\hspace{\parindent}%
There exist many ways in the literature
to algebraically compute
(one-loop) anomalies. All of them are
essentially based
in testing the response of the
--suitably regulated-- partition
function of the model under the
(infinitesimal version of
the) symmetry transformation under study.
Departures from unity of the jacobian
arising upon this change which
can not be absorved by suitable
counterterms reflect then the
presence of anomalies in the model.

The so-called ``nonlocal regularization''
method, recently introduced in
\cite{emkw,j95}, fits perfectly well
in this philosophy. Indeed, this
approach proceeds by constructing
from the original action $S(\Phi^A)$
and symmetry transformations
$\delta\Phi_A$ of the model a regulated
action $S_\Lambda(\Phi^A)$,
invariant under a ``regulated''
version of the original symmetry,
$\delta_\Lambda\Phi_A$, where $\Lambda$
stands for a cut-off or regulating parameter.
Such invariant action,
exponentiated afterwards in the path integral,
generates then a modified
set of Feynman rules and propagators that
yield finite Feynman integrals
for finite values of the cut-off at all
loop levels, and thus a finite
partition function.

For our purposes, there are two main
advantages of this approach
relative to other
``standard'' regularization methods.
First of all,
the nonlocally regularized action
$S_\Lambda(\Phi^A)$
can just be seen as a ``smooth''
deformation of the original one such
that its main features
(dimensionality, field content, symmetries...)
remain unaltered.
Therefore, when dealing with supersymmetric
theories, in
particular, superspace computations at
regulated level can be performed in
exactly the same way as in the original theory.
Second, and on top of that,
the invariance of $S_\Lambda$ under $\delta_\Lambda$
directly relates
potential one--loop anomalies to the
finite part of the functional trace
--now completely regulated-- of the
jacobian matrix, namely%
\footnote{De Witt notation is assumed
throughout the paper whenever
capital indices $A,B,\ldots$ are used.
These indices
indicate the different fields, their components,
{\em and} the
space-time point on which they depend (unless it
is explicitly displayed).
In this way, a summation over $A$ includes not
only discrete summations,
but also integration over (super)space-time.
The derivatives are left and right functional derivatives.}
\begin{equation}
  {\cal A}= \left[(-1)^{A}
  \frac{\partial_r (\delta_\Lambda\Phi_A)}
  {\partial\Phi_A}\right],
\label{gen anom}
\end{equation}
where $(-1)^{A}\equiv (-1)^{|\Phi_A|}$
stands for the Grassmann parity of the field $\Phi_A$.
In view of these facts,
nonlocal regularization appears thus as an
excellent candidate to implement our programme.

In what follows, we briefly
summarize the construction of the
nonlocal action $S_\Lambda$ and of
its symmetries $\delta_\Lambda$, as
well as the specific form of the
anomaly (\ref{gen anom}), along the lines
of refs.\,\cite{emkw,j95}, implementing
afterwards the standard superspace
techniques in this framework.

\subsection{Basics on Nonlocal Regularization}
\label{nlr}

\hspace{\parindent}%
Consider a theory defined by a classical
action ${\cal S}(\Phi^A)$,
which admits a sensible perturbative
decomposition into free and
interacting parts
\begin{equation}
  {\cal S}(\Phi)= F(\Phi)+ I(\Phi),
  \quad\quad \mbox{\rm with}\quad\quad
   F(\Phi)=\frac12\Phi^A F_A{}^B \Phi_B.
\label{original action}
\end{equation}
Introduce now a field independent operator
$(T^{-1})_A{}^B$ such that
a second order derivative `regulator'
$R_A{}^B$ arises through the combination
$$
   R_A{}^B=(T^{-1})_A{}^C F_C{}^B,
$$
and construct from this object the
so-called smearing operator,
${\cal{\varepsilon}}_A{}^B$, and shadow
kinetic operator, $({\cal O}^{-1})_A{}^B$
\begin{equation}
   {\cal{\varepsilon}}_A{}^B=
   \exp\left(\frac{R_A{}^B}{2\Lambda^2}\right),
\label{smearing op}
\end{equation}
\begin{equation}
   ({\cal O}^{-1})_A{}^B=
   T_A{}^C\, \int^1_0\frac{{\rm d} t}{\Lambda^2}\,
   \exp\left(t\frac{R_C{}^B}{\Lambda^2}\right).
\label{shadow kinetic op}
\end{equation}

To each original field $\Phi_A$
it is now associated an auxiliary, or
`shadow', field $\Psi_A$ with the
same statistics. Both sets of
fields are then coupled by means
of the auxiliary action
\begin{equation}
  \tilde{\cal S}(\Phi,\Psi)=
  F(\hat\Phi)-A(\Psi)+  I(\Phi+\Psi),
\label{auxiliary action}
\end{equation}
with $A(\Psi)$, the kinetic term
for the auxiliary fields,
constructed with the help of
(\ref{shadow kinetic op}) as
$$
  A(\Psi)=\frac12\Psi^A ({\cal O}^{-1})_A{}^B \Psi_B,
$$
and where the ``smeared'' fields $\hat\Phi_A$
appearing in the free part of
the auxiliary action (\ref{auxiliary action}) are defined,
using (\ref{smearing op}),
by $\hat\Phi_A\equiv({\cal{\varepsilon}}^{-1})_A{}^B\Phi_B$.

The perturbative theory described
by (\ref{auxiliary action}), when only
external $\Phi$ lines are considered,
is then seen to describe the same
theory as the original action (\ref{original action}).
However, the special
form of propagators and couplings in (\ref{auxiliary action})
lead the
loops formed with shadow propagators to isolate the divergent
parts of the
original diagrams. As a consequence,
dropping out these loop contributions,
i.e., the quantum fluctuations of the
shadow fields by hand, regularizes
the theory. Such {\it ad hoc}
procedure may however be simply
implemented by putting the auxiliary
fields $\Psi$ classically on--shell.
The classical shadow field equations of motion
\begin{equation}
  \frac{\partial_r \tilde{\cal S}(\Phi,\Psi)}
  {\partial \Psi_A}=0
  \quad\Rightarrow\quad
  \Psi^A= \left(\frac{\partial_r I}{\Phi_B}
  (\Phi+\Psi)\right){\cal O}_B{}^A,
\label{shadow eqs motion}
\end{equation}
should then be solved, in general,
in a perturbative fashion
and its solution $\Psi_0(\Phi)$
substituted in the
auxiliary action (\ref{auxiliary action}).
The result of this process is the
nonlocalized action to be used
in regularized perturbative computations
\begin{equation}
  {\cal S}_{\Lambda}(\Phi)\equiv
  \tilde{\cal S}(\Phi,\Psi_0(\Phi)).
\label{nonlocal action}
\end{equation}

Moreover, as mentioned above,
the nonlocalization procedure
just presented has the merit
of preserving at tree level a distorted
version of any of the original
continuous symmetries of the theory.
Indeed, assume the original action
(\ref{original action}) be invariant
under the infinitesimal transformation
$$
   \delta\Phi_A= R_A(\Phi).
$$
Then, the auxiliary action
(\ref{auxiliary action})
is seen to be invariant under
the auxiliary infinitesimal transformations
$$
   \tilde\delta\Phi_A=
   ({\cal{\varepsilon}}^2)_A{}^B R_B(\Phi+\Psi),
   \quad\quad
   \tilde\delta\Psi_A=
   (1-{\cal{\varepsilon}}^2)_A{}^B R_B(\Phi+\Psi),
$$
while the nonlocally regulated action
${\cal S}_{\Lambda}(\Phi)$
(\ref{nonlocal action}) becomes invariant under
$$
   \delta_{\Lambda}\Phi_A=
   ({\cal{\varepsilon}}^2)_A{}^B R_B(\Phi+\Psi_0(\Phi)),
$$
with $\Psi_0(\Phi)$ the solution of
(\ref{shadow eqs motion}).
In this way, an extensive use of the chain rule
allows to determine a closed form for
the anomaly (\ref{gen anom}) in
terms of propagators and vertices
of the original theory as
\begin{equation}
  {\cal A}= \left[(-1)^{A}
  ({\cal{\varepsilon}}^2)_A{}^B\,
  J_B{}^C \,(\delta_\Lambda)_C{}^A\right],
  \quad\quad \mbox{where}\quad\quad
  J_A{}^B=
  \frac{\partial_r R_A}{\partial\Phi_B},
\label{gen anom 2}
\end{equation}
and with the regulated identity
$(\delta_\Lambda)_A{}^B$ defined by
$$
  (\delta_\Lambda)_A{}^B\equiv
  \left(\delta_A{}^B- {\cal O}_A{}^C I_C{}^B\right)^{-1}=
   \delta_A{}^B+\sum_{n\geq 1}
  \left({\cal O}_A{}^C I_C{}^B\right)^{n},
$$
in terms of the functional hessian
of the original interaction in
(\ref{original action})
\begin{equation}
  I_A{}^B=
  \frac{\partial_l\partial_r I}
  {\partial\Phi^A\partial\Phi_B}.
\label{hessian int}
\end{equation}
The proof of these statements is
straightforward and can be found in the
original references \cite{emkw,j95},
to which we refer the reader for
further details.

\subsection{Implementation of superspace techniques}
\label{impl}

\hspace{\parindent}%
The nonlocal regularization procedure outlined above
applies of course to all kinds of perturbative models,
including
supersymmetric ones. Now, it is well-known
that in supersymmetric theories
perturbative calculations
can often be considerably simplified
by means of superspace techniques
due to the cancellation of terms
caused by supersymmetry.
It is therefore natural to look for
a way to implement these techniques in the
nonlocal regularization procedure.
An obvious idea is
to replace ordinary fields by superfields.
However one faces
immediately the following related difficulties:
how should one
define functional derivatives w.r.t. arbitrary
(constrained) superfields
and integrations over their `superspace coordinates'?
These two problems
appear to make the simple substitution
`fields $\rightarrow$ superfields'
impossible except in very special cases
where one deals only with
particular superfields such as unconstrained or
chiral ones. Thus in general we cannot
simply take the $\Phi$'s of the previous
subsections to be superfields.

Fortunately this is not necessary
at all since superspace techniques
are of course not restricted to
true superfields%
\footnote{See appendix \ref{disc}
for a discussion of the concept of
superfield.}. In fact, we will show
now that they apply
also to ``constituents'' of superfields such as
\begin{equation}
\varphi(x,\bar \theta)=
a(x)+\bar \theta_{\dot{\alpha}}
b^{\dot{\alpha}}(x)+
{\mbox{\small{$\frac12$}}}\bar \theta^2 c(x)\ ,
\label{F1}
\end{equation}
provided $a,b,c$ are elementary fields. Namely then we
can {\em define} functional derivatives%
\footnote{For definiteness
all formulae are written for
left-derivatives in this subsection.}
w.r.t.\ $\varphi$ simply through
\begin{equation}
\frac {\partial}{\partial \varphi(x,\bar \theta)}=
-\frac {\partial}{\partial c(x)}
+\bar \theta^{\dot{\alpha}}
\frac {\partial}{\partial b^{{\dot{\alpha}}}(x)}
-{\mbox{\small{$\frac12$}}}
\bar \theta^2\frac {\partial}{\partial a(x)}\ ,
\label{im1}
\end{equation}
which results in
$$
\frac {\partial \varphi(x,\bar \theta)}
{\partial \varphi(x',\bar \theta')}
=\delta^2(\bar \theta-\bar \theta')
\delta^4(x-x')\equiv\delta^6(\bar z-\bar z').
$$
Summation over their indices in de Witt's
condensed notation includes then
simply an integration
$\int d^6\bar z\equiv\int d^4x d^2\bar \theta$.

Alternatively we can (and will) use
instead of $\varphi$ the quantity
\begin{equation}
\Phi(z)=\exp(-i\theta\partial \bar \theta)\,
\varphi(x,\bar \theta)\ ,
\label{im3}
\end{equation}
which is antichiral in the sense that
\begin{equation}
{\cal D}_\alpha\Phi=0,
\label{im4}
\end{equation}
where the standard covariant
derivatives are defined as
\begin{equation}
{\cal D}_\alpha=\frac{\partial }
{\partial \theta^\alpha}
+i\bar \theta^{\dot{\alpha}}
\partial _{\alpha{\dot{\alpha}}}\ ,\quad
\bar {\cal D}_{\dot{\alpha}}=
-\frac{\partial }{\partial \bar \theta^{\dot{\alpha}}}
-i\theta^\alpha\partial _{\alpha{\dot{\alpha}}}\ .
\label{sf3}
\end{equation}
However $\Phi$ is not in general
a superfield (see appendix \ref{disc})
i.e.\ (\ref{im4}) does not reflect
the transformation properties
of $\Phi$.
The functional derivative w.r.t.\
$\Phi$ is then defined
by means of (\ref{im1}) according to
$$
\frac {\partial}{\partial \Phi(z)}=
\exp(-i\theta\partial \bar \theta)
\frac {\partial}{\partial \varphi(x,\bar \theta)}\ .
$$
This results in
\begin{equation}
\frac {\partial \Phi(z)}{\partial \Phi(z')}=
{\mbox{\small{$\frac12$}}} {\cal D}^2\delta^8(z-z')\ ,
\label{im6}\end{equation}
due to the identity
$$\exp(-i\theta\partial \bar \theta+
i\theta'\partial \bar \theta')\delta^6(\bar z-\bar z')=
{\mbox{\small{$\frac12$}}}{\cal D}^2\delta^8(z-z').
$$
Formula (\ref{im6}) can indeed be found in many textbooks
on supersymmetry for functional
derivatives w.r.t.\ to antichiral superfields -- we just
extend it to constituents
of superfields satisfying (\ref{im4}). Due
to the presence of the
antichiral projector
${\mbox{\small{$\frac12$}}}{\cal D}^2$ in (\ref{im6}),
summation over the indices of these constituents does not
involve the integration $\int d^8z$ but again only
an integration $\int d^6\bar z$. Analogous formulae
hold of course for functional right-derivatives and
chiral quantities.

We conclude that we can use quantities like
(\ref{F1}) or (\ref{im3}) in nonlocal regularization
instead of ordinary fields.
This remains true even if it is
impossible to combine all the elementary fields in
such quantities -- the remaining elementary fields may
be treated as usual, i.e.\ one can use
quantities (\ref{F1}) or (\ref{im3}) and ordinary fields
simultaneously if necessary. The only thing
one has to keep in mind when
dealing with such constituents is that
operators such as (\ref{sf3})
or the usual generators of supersymmetry
transformations
\begin{equation}
\nabla_\alpha=\frac{\partial }{\partial \theta^\alpha}
-i\bar \theta^{\dot{\alpha}}
\partial _{\alpha{\dot{\alpha}}}\ ,\quad
\bar \nabla_{\dot{\alpha}}=
-\frac{\partial }{\partial \bar \theta^{\dot{\alpha}}}
+i\theta^\alpha\partial _{\alpha{\dot{\alpha}}},
\label{sf1}
\end{equation}
do not have
the same interpretation on constituents of superfields
as on superfields themselves: in particular
the operators (\ref{sf1}) do not represent the supersymmetry
transformations anymore on all of the constituent fields.

\section{The model}\label{model}

\subsection{Multiplet and supersymmetry transformations}
\label{mult}

\hspace{\parindent}%
The four dimensional toy model we are going
to use contains only a
supersymmetry multiplet considered in section 7 of
\cite{rigid}.
This multiplet consists of complex Weyl spinors
$\chi$, $\psi$ and $\eta$, a complex vector
field $V$, and two complex
scalar fields $A$ and
$F$. On these fields the abstract
supersymmetry algebra
\begin{eqnarray}
& & [P_a,P_b]=[P_a,Q_\alpha]=
[P_a,\bar Q_{\dot{\alpha}}]=0,
\nonumber\\
& & \{Q_\alpha,Q_\beta\}=
\{\bar Q_{\dot{\alpha}},
\bar Q_{\dot{\beta}}\}=0,\quad
    \{Q_\alpha,\bar Q_{\dot{\alpha}}\}=
    -2i \sigma^a{}_{\alpha{\dot{\alpha}}}P_a,
\label{susyalg}
\end{eqnarray}
is represented with
$(P_a,Q_\alpha,\bar Q_{\dot{\alpha}})
\equiv(\partial_a,D_\alpha,\bar D_{\dot{\alpha}})$
according to table 1
(using $X_{\alpha{\dot{\alpha}}}=
\sigma^a{}_{\alpha{\dot{\alpha}}}X_a$).
\[ \begin{array}{c||c|c|c|c|c|c}
\phi &\chi_\beta  &A&V_{\beta{\dot{\beta}}}
&\bar \psi_{\dot{\beta}}&\eta_\beta&F\\
\hline
\rule{0em}{3ex}
D_\alpha\phi &\varepsilon_{\beta\alpha}\, A &0&
-2i\partial_{\alpha{\dot{\beta}}}\,
\chi_\beta+\varepsilon_{\alpha\beta}\,
\bar \psi_{\dot{\beta}}
&-2i\partial_{\alpha{\dot{\beta}}}\, A&
2i\partial_{\alpha{\dot{\alpha}}}{V_\beta}^{\dot{\alpha}}
-\varepsilon_{\alpha\beta}\, F
&2i\partial_{\alpha{\dot{\alpha}}}\bar \psi^{\dot{\alpha}}\\
\bar D_{\dot{\alpha}}\phi
&V_{\beta{\dot{\alpha}}}&\bar \psi_{\dot{\alpha}}&
\varepsilon_{{\dot{\alpha}}{\dot{\beta}}}\, \eta_\beta
&\varepsilon_{{\dot{\alpha}}{\dot{\beta}}}\, F&0&0\\
\mbox{dim} (\phi) & 1/2 & 1 & 1 & 3/2 & 3/2 & 2 \\
\multicolumn{7}{c}{}\\
\multicolumn{7}{c}{\mbox{Table 1}}
\end{array}
\]
The assignment of the dimensions (dim)
to the fields in table 1 follows
from the choice dim$(\chi)$=1/2,
which will be the
power counting dimension of $\chi$, and
from the standard convention dim$(D_\alpha)$=
dim$(\bar D_{\dot{\alpha}})$=1/2,
dim$(\partial_a)$=1. Supersymmetry transformations,
$\delta_{susy}$, of the
fields in table 1 are then defined
according to the relation
\begin{equation}
\delta_{susy}=\epsilon^\alpha D_\alpha+
\bar \epsilon_{\dot{\alpha}} \bar D^{\dot{\alpha}}
\equiv \epsilon^{\underline{\alpha}}
D_{\underline{\alpha}}\ ,
\label{susytrafo}
\end{equation}
where the parameters $\epsilon^\alpha$
are constant anticommuting spinors.

The supersymmetry multiplet and
transformation laws of table 1 can
also be formulated in superspace
(c.f. appendix \ref{disc}) which will be
useful within the computation of
the anomaly coefficients.
However, for the reasons we have just explained,
we will apply a somewhat unconventional
approach involving
not only true superfields but
also special constituents of them,
which will be introduced and discussed in the following.

The fundamental (`defining') superfield of
the multiplet of table 1 is
\begin{equation}
G^\alpha=\exp(\theta D+\bar \theta \bar D)\,
\chi^\alpha=H^\alpha+\theta^\alpha K\ ,
\label{superfield}
\end{equation}
with
\begin{eqnarray}
& &H^\alpha=\exp (-i\theta\partial \bar \theta)\,
h^\alpha,\quad
K=\exp (-i\theta\partial \bar \theta)\, k,\label{sf6}\\
& & h^\alpha=\exp(\bar \theta \bar D)\, \chi^\alpha=
\chi^\alpha+\bar \theta_{\dot{\alpha}}
V^{{\dot{\alpha}}\alpha}+{\mbox{\small{$\frac12$}}}
\bar \theta^2\eta^\alpha ,
\label{hdef}\\
& & k=\exp(\bar \theta \bar D)\,
A=A+\bar \theta_{\dot{\alpha}}
\bar \psi^{\dot{\alpha}}+{\mbox{\small{$\frac12$}}}
\bar \theta^2 F\ ,
\label{ldef}
\end{eqnarray}
where we used the identity (\ref{useful}),
table 1 and the notation
$\theta\partial \bar \theta=
\theta^\alpha\partial _{\alpha{\dot{\alpha}}}
\bar \theta^{\dot{\alpha}}$,
$\theta^2=\theta^\alpha \theta_\alpha$ and
$\bar \theta^2=\bar \theta_{\dot{\alpha}}
\bar \theta^{\dot{\alpha}}$.
The split of $G^\alpha$ into
the constituents $H^\alpha$ and
$K$ will be useful later on,
in particular since the latter are
`antichiral' in the sense that%
\footnote{Throughout the paper
superfields or
constituents thereof are called
antichiral if they satisfy eq.\,(\ref{sf6a})
and (functions of) elementary fields
and their derivatives are
called antichiral if they fulfill $D_\alpha \phi=0$.}
\begin{equation}
{\cal D}_\alpha H_\beta={\cal D}_\alpha K=0\ ,
\label{sf6a}
\end{equation}
whereas $G^\alpha$ itself satisfies the `constraint'
\begin{equation}
{\cal D}_{(\alpha}G_{\beta)}=0.
\label{constraint}
\end{equation}
It is important to realize and keep in mind
that $H_\alpha$ is {\em not} a superfield since it
does not satisfy the first identity
(\ref{sf2}). Rather, its supersymmetry transformations
are given by
\begin{equation}
D_\alpha H_\beta=\nabla_\alpha H_\beta
+\varepsilon_{\beta\alpha} K\ ,\quad
\bar D_{\dot{\alpha}} H_\beta =
\bar \nabla_{\dot{\alpha}} H_\beta \ .
\label{sf7}
\end{equation}
In contrast, $K$ is a true superfield
and thus satisfies (\ref{sf2}),
\begin{equation}
K={\mbox{\small{$\frac12$}}}{\cal D}_\alpha G^\alpha\ ,
\quad
 D_\alpha K=\nabla_\alpha K,\quad \bar D_{\dot{\alpha}} K
 = \bar \nabla_{\dot{\alpha}} K.
\label{sf8}\end{equation}

We remark that
the supersymmetry multiplet of table 1 can be truncated
(consistently with the supersymmetry algebra)
in two ways, by setting to zero either
all the fields $\chi,V,\eta$
or all the fields $A,\psi,F$. One would then be left with
standard antichiral supersymmetry multiplets given by
$(A,\bar \psi,F)$ and $(\chi,V,\eta)$ respectively,
corresponding
to $K$ and $H^\alpha$ respectively.
Hence, the supersymmetry multiplet of
table 1 may be regarded as a
nontrivial merger of these two multiplets.
Alternatively, one can regard it itself
as the truncation of a full complex vector multiplet
corresponding to an unconstrained
complex scalar superfield.

\subsection{Action}

\hspace{\parindent}%
Using the techniques of \cite{rigid} one can prove
that the most general real action
for the supersymmetry multiplet of table 1 which is
a) polynomial in the elementary
fields and their derivatives,
b) constructible out of field
monomials of dimension $\leq 4$ (with
dimensions as in table 1), c) Poincar\'e invariant and d)
invariant (up to surface terms)
under the supersymmetry transformations
 $D_\alpha$ and $\bar D_{\dot{\alpha}}$
 given in table 1,
can be written, up to surface terms,
in terms of superspace integrals in the form
\begin{eqnarray}
S&=&\int d^4x\, (L_1+L_2+L_3+L_4),
\label{a1}\\
L_1&=&\int d^2\bar \theta\,
\left\{\mu^2K+ \mbox{\rm c.c.}\right\}\ ,
\label{a2}\\
L_2&=&\int d^4\theta\,
\{ia_1G\partial \bar G
+a_2K\bar K+({\mbox{\small{$\frac14$}}}a_3G\bar {\cal D}^2G
+{\mbox{\small{$\frac12$}}}mGG+\mbox{\rm c.c.})\}\ ,
\label{a3}\\
L_3&=&\int d^4\theta\,
\left\{({\mbox{\small{$\frac12$}}}b_1GG\bar K
+{\mbox{\small{$\frac12$}}}b_2GGK)
+\mbox{\rm c.c.}\right\}\ ,
\label{a4}\\
L_4&=&\int d^4\theta\, {\mbox{\small{$\frac14$}}}b_3 GG\,
\bar G\bar G\ ,
\label{a5}
\end{eqnarray}
where $G^\alpha$ and $K$ are the superfields
given in (\ref{superfield}) and (\ref{sf6});
$\mu^2,a_3,m,b_1,b_2$ are complex parameters
and $a_1,a_2,b_3$
are real parameters. The action is spelled
out explicitly
in appendix \ref{appact}.

Some special features of this general
action merit now special
consideration. First of all,
the terms in (\ref{a2}-\ref{a5})
corresponding to the parameters
$\mu^2,m,b_2$ give rise to a
superpotential $(\mu^2 K-m K^2-b_2 K^3)$
for the antichiral multiplet
$(A,\bar \psi,F)$ since one has
\begin{equation}
\int d^2\bar \theta\,  \mu^2 K+
{\mbox{\small{$\frac12$}}}\int d^4\theta\,
(m G G+b_2 G G K)
\cong \int d^2\bar \theta\, (\mu^2 K-m K^2-b_2 K^3),
\label{a6}
\end{equation}
where $\cong$ denotes
equality up to a total derivative.
Expression (\ref{a6}) together
with the kinetic term corresponding to the
parameter $a_2$ constitute thus nothing but the
familiar action of a
Wess-Zumino model for the fields $A,\psi,F$ making up
the (anti)chiral superfields $K$, $\bar K$.
The other terms in the action
involve also the fields $\chi,V,\eta$
and in particular couple them to $A,\psi,F$.

For simplicity we will later not work with
the above general action but restrict
ourselves to the simpler action
\begin{equation}
\int d^8z \, (ia_1G\partial \bar G
+a_2K\bar K+{\mbox{\small{$\frac12$}}}b_1GG\bar K
+{\mbox{\small{$\frac12$}}}\bar b_1\bar G\bar GK),
\label{simpleact}
\end{equation}
i.e.\ we will set to zero
the Wess-Zumino superpotential (\ref{a6}) as well as
the coefficients $a_3$ and $b_3$.
Furthermore we will assume
\begin{equation}
a_1\neq 0,\quad a_1+a_2\neq 0\ ,
\label{sing}\end{equation}
since otherwise (\ref{simpleact})
does not give well-defined propagators for all the fields.
$a_1\neq 0$ is imposed since otherwise
the kinetic terms of (\ref{simpleact}) reduce to those of
the Wess-Zumino model for $A,\psi,F$ and the remaining
fields would not propagate. $a_1+a_2\neq 0$ warrants
that (\ref{simpleact}) has no gauge invariance.

\subsection{Candidate anomalies}\label{anos}

\hspace{\parindent}%
By standard arguments, analogous to those used in
\cite{wz} and applied to the
vertex functional (effective action),
one concludes from the (classical) supersymmetry algebra
(\ref{susyalg}) that at lowest order in $\hbar$
supersymmetry anomalies must
satisfy the consistency conditions
\begin{equation}
D_{(\alpha} \Delta_{\beta)}
=\bar D_{({\dot{\alpha}}} \Delta_{{\dot{\beta}})}
=D_\alpha \Delta_{\dot{\alpha}}
+\bar D_{\dot{\alpha}} \Delta_\alpha=0\ ,
\label{can1}\end{equation}
where the contributions
$\Delta_\alpha$ and $\Delta_{\dot{\alpha}}$ to such
anomaly are local functionals of the fields.
Furthermore one can assume
\begin{equation}
\Delta_\alpha\neq D_\alpha \Gamma_0,\quad
\Delta_{\dot{\alpha}}\neq \bar D_{\dot{\alpha}} \Gamma_0\ ,
\label{can2}
\end{equation}
for any local functional $\Gamma_0$
of the fields since otherwise
the anomaly can be removed through a
local counterterm, at least up to
terms of higher order in $\hbar$.

The consistency condition (\ref{can1})
and the non-triviality
condition (\ref{can2})
are most efficiently formulated and
analysed using cohomological
techniques. To that end one introduces a
`BRST'-operator $s$ corresponding
to the algebra (\ref{susyalg})
$$
s=\xi^\alpha D_\alpha+\bar \xi^{\dot{\alpha}}
\bar D_{\dot{\alpha}}+C^a\partial_a+2i\xi\sigma^a\bar \xi
\frac \partial {\partial  C^a}\ ,
$$
where $\xi^\alpha$ are constant commuting
supersymmetry ghosts and $C^a$
are constant anticommuting translation
ghosts ($D_\alpha$ and
$\bar D_{\dot{\alpha}}$ vanish on the ghosts).
$s$ is nilpotent and allows
to reformulate (\ref{can1}) and
(\ref{can2}) through
\begin{equation}
s\, \Delta=0,\quad \Delta\neq s\, \Gamma_0\ ,
\label{cc}
\end{equation}
with
$$
\Delta=\xi^\alpha\Delta_\alpha
+\bar \xi^{\dot{\alpha}}\Delta_{\dot{\alpha}}\ .
$$
In (\ref{can1}) and (\ref{cc})
it is understoood that the operators
($D_{\underline{\alpha}}$ resp.\ $s$)
act on the integrands of the $\Delta$'s
and $\Gamma_0$ and,
in general, equalities need to hold
only on--shell
(up to surface terms).

For the model in question two complex
solutions of (\ref{cc}) have been
given in section 7 of \cite{rigid}:
\begin{equation}
\Delta_1=\xi^\alpha\int d^4x \bar D^2\chi_\alpha
=-2\xi^\alpha\int d^4x \,\eta_\alpha\ ,\quad
\Delta_2=\xi^\alpha\int d^4x
\bar D^2(\chi_\alpha\bar \psi'\bar \psi')\ ,
\label{can5}\end{equation}
where $\bar \psi'$ is the combination
\begin{equation}
\bar \psi'_{\dot{\alpha}}=
\bar \psi_{\dot{\alpha}}+
2i\partial_{\alpha{\dot{\alpha}}}\chi^\alpha\ .
\label{prime}
\end{equation}
The explicit form of $\Delta_2$
is given in appendix \ref{appact}.
We note that both $\Delta_1$ and $\Delta_2$ give in fact rise to
two independent real solutions of (\ref{cc}),
given by their  real and imaginary part respectively.

Using the methods of \cite{rigid} and extending them
to the on-shell problem%
\footnote{This is done efficiently
by introducing antifields \`a la \cite{bv}.}
one can prove that, up to trivial
solutions of the form $s\Gamma_0$ and surface terms,
the functionals (\ref{can5}) and their
complex conjugates are indeed the only
inequivalent solutions to (\ref{cc}) in our model
which have the correct Lorentz transformation properties
and are polynomials in all the fields
and their derivatives with dim$(\Delta)\leq 4$
(using dim$(\xi)=-1/2$).

It is evident that both functionals
(\ref{can5}) indeed solve the
first condition (\ref{cc}), using the fact that
$\bar \psi'$ is antichiral, i.e.
$$
D_\alpha\bar \psi'_{\dot{\alpha}}=0.
$$
Furthermore $\Delta_1$ and
$\Delta_2$ are cohomologically nontrivial, i.e.
there is no local functional $\Gamma_0$ of the fields such that
$s\Gamma_0$ equals $\Delta_1$ or $\Delta_2$ on-shell
modulo a surface term. This can be verified straightforwardly by
an explicit inspection of all the relevant candidates for $\Gamma_0$.
In fact there are
only finitely many such candidates as only functionals
need to be considered which have the same dimension as
the respective $\Delta$ (1 resp. 4) and which are
Lorentz-invariant, thanks to the properties of $s$.

Without going into details we remark that the presence of
candidate supersymmetry anomalies in our model is due to the fact that
the representation of the supersymmetry algebra  given in 
table 1 of section \ref{mult} does not have "QDS-structure"
in the terminology of \cite{rigid}, in contrast to more
standard representations of supersymmetry. 
Furthermore we note that the non-QDS-property itself
can be traced back to the `constraint' (\ref{constraint}).

Finally we add two comments concerning the
consistency condition for supersymmetry
anomalies in general
and its solutions $\Delta_1$ and $\Delta_2$:

a) In superspace notation $\Delta_1$ and $\Delta_2$ read
\begin{equation}
\Delta_1=-\xi^\alpha\int d^8z\,  \theta^2\,  G_\alpha\ ,\quad
\Delta_2=- \xi^\alpha \int d^8z\,
\theta^2\, G_\alpha\, \bar \Psi'\bar \Psi'\ ,
\label{cand1}
\end{equation}
with $G_\alpha$ as in (\ref{superfield}) and 
$\bar \Psi'$ being the antichiral superfield
whose lowest component field is
$\bar \psi'$ (\ref{prime})
$$
\bar \Psi'_{\dot{\alpha}}=
\exp(\theta D+\bar \theta\bar D)\,
\bar \psi'_{\dot{\alpha}}
=\bar {\cal D}_{\dot{\alpha}} K+
2i\partial_{\alpha{\dot{\alpha}}}G^\alpha\ .
$$
The presence of
$\theta^2$ in the integrands in (\ref{cand1})
indicates that $\Delta_1$ and $\Delta_2$
cannot be written as
superspace integrals $\int d^8z$
(or $\int d^6\bar z$) over
true (antichiral) superfields. 
%We will see
%in section \ref{comp} that the
%functionals arising in the
%perturbative computation of the
%anomaly coefficients
%share this property.
This shows that in general it would
be misleading to formulate
the consistency conditions (\ref{can1}) resp.\ (\ref{cc})
in terms of
the operators $\nabla_{\underline{\alpha}}$
defined in (\ref{sf1}) instead
of the $D_{\underline{\alpha}}$
(recall that the $\nabla$'s represent
the supersymmetry transformations
only on true superfields).

b) The dimensions of $\Delta_1$
and $\Delta_2$ indicate that
they would play different roles
if they would occur in
the (anomalous) jacobian of
supersymmetry transformations:
$\Delta_1$ has dimension 1 and thus
would eventually arise
as a {\em divergent} contribution to
that jacobian, in contrast to
$\Delta_2$ which has canonical dimension 4
and is interpreted as a genuine potential anomaly.

\section{Computation of the anomaly coefficients}
\label{comp}

\hspace{\parindent}%
Let us finally pass to investigate
the actual presence of the candidate
anomalies (\ref{can5})
in our toy model by applying
expression (\ref{gen anom 2})
of the nonlocally regularized form of the
anomaly to it. For the sake of simplicity,
to illustrate the procedure and
results we restrict ourselves to the simple
version (\ref{simpleact}) of the
general action (\ref{a1}).

The structure of the superfield (\ref{superfield})
and the previous considerations immediately
suggest to work with its
`(anti)chiral' constituents (\ref{sf6})
and use as basis to express
matrix-like operators
\begin{eqnarray*}
& &\Phi^A\equiv (\Phi^a, \Phi_{\bar a})\equiv
(H^\alpha,K;\bar H_{\dot{\alpha}},\bar  K),\quad\quad
\Phi_A\equiv \left(\begin{array}{c} \Phi_a \\
\Phi^{\bar a}\end{array}\right)\equiv
\left(\begin{array}{c}H_\alpha\\ K \\
\bar H^{\dot{\alpha}} \\ \bar  K
\end{array}\right),
\end{eqnarray*}
where latin indices express compactly
antichiral $(a)$ and chiral
$(\bar a)$ components.
In terms of these (anti)chiral components, t
he action (\ref{simpleact})
reads then
\begin{eqnarray}
     S&=&\int{\rm d}^8 z\left\{
     i a_1 (H^\alpha
     +\theta^\alpha K) \partial_{\alpha{\dot{\alpha}}}
     (\bar H^{\dot{\alpha}}
     +\bar\theta^{\dot{\alpha}} \bar K)
     + a_2 K\bar K\right.
\nonumber\\
     &&\left.+\left[{\mbox{\small{$\frac12$}}}b_1
     \left(H^\alpha H_\alpha +
     2 H^\alpha\theta_\alpha K
     +\theta^2 K^2\right) \bar K
     + \mbox{(c.c.)} \right]\right\}.
\label{compaction}
\end{eqnarray}

As pointed out in subsection
\ref{impl} (and in many textbooks),
the constrained character of these
(anti)chiral components requires
some reinterpretation of their superspace integration
and functional differentiation rules.
First of all, the functional
derivative rules for (anti)chiral
fields (\ref{im6}), now reading
$$
  \frac {\partial \Phi^a}{\partial \Phi^b}=
  \frac {\partial \Phi_b}{\partial \Phi_a}=
  {\mbox{\small{$\frac12$}}} {\cal D}^2 \,\delta^a_b,
$$
where $\delta^a_b$ encodes,
according to the compact notation we are using,
a discrete identity as well as
the $8$-dimensional delta function
$\delta^8(z-z')$ in superspace,
express nothing but the fact that
(anti)chiral fields
and operators obtained from functional
differentiation with respect
to them naturally live in six dimensional
superspace. This fact is
conveniently expressed by introducing the
projector in the space of
antichiral-chiral superfields $(P_q)_A{}^B$
$$
  (P_q)_A{}^B=
  \left( \begin{array}{cc}
  (P_q)_a{}^b & 0\\
  0& (P_q)^{\bar a}{}_{\bar b}
  \end{array}\right)=
  \left( \begin{array}{cc}
  {\mbox{\small{$\frac12$}}}{\cal D}^2\,
  \delta_a{}^b& 0\\
  0& {\mbox{\small{$\frac12$}}} \bar{\cal D}^2
  \, \delta^{\bar a}{}_{\bar b}
  \end{array}\right),
$$
verifying%
\footnote{Recall that matrix multiplication
among projectors $P$
must be performed using an integration in the
corresponding {\em six} dimensional superspaces,
i.e.\, either
$\int {\rm d}^6 \bar z$ or $\int {\rm d}^6 z$.}
\begin{equation}
  (P_q)_a{}^c (P_q)_c{}^b=
  \int{\rm d}^6 \bar z''\,{\mbox{\small{$\frac12$}}}
  {\cal D}^2_z \,\delta^8(z-z'')\,
  {\mbox{\small{$\frac12$}}}{\cal D}^2_{z''}\,
  \delta^8(z''-z')=
  {\mbox{\small{$\frac12$}}}{\cal D}^2_z \,
  \delta^8(z-z')= (P_q)_a{}^b,
\label{idempotency}
\end{equation}
and an analogous relation for the chiral
sector. `(Anti)chiral' kernels
will thus be typically expressed,
in compact notation, as
$$
  M_A{}^B = (P_q)_A{}^C {\cal M}_C{}^D (P_q)_D{}^B
  \equiv (P_q {\cal M} P_q)_A{}^B,
$$
so that super matrix multiplication
will then yield, according to
(\ref{idempotency})
$$
  M_A{}^C N_C{}^B=
  (P_q {\cal M} P_q)_A{}^C(P_q {\cal N} P_q)_C{}^B=
  (P_q {\cal M} P_q {\cal N} P_q)_A{}^B.
$$

The nonlocal regularization of the
model (\ref{compaction})
requires now the identification of
the basic quantities involved
in the computation, namely
the jacobian (\ref{gen anom 2})
of the original transformation,
the hessian of the interaction
(\ref{hessian int})
and the regulating objects
related to the kinetic operator
(\ref{original action}).
The jacobian of the original
transformation (\ref{susytrafo}) adopts in the
above basis, according to
eqs.\,(\ref{sf7}), (\ref{sf8}), the form
\begin{equation}
  J_A{}^B=
  \frac{\partial_r
  \left(\delta_{susy}\Phi_A\right)}{\partial\Phi_B}=
  ({\cal J} P_q)_A{}^B=
  \left(\begin{array}{cc}
   {\mbox{\small{$\frac12$}}}
   {\cal D}^2 \,{\cal J}_a{}^b & 0 \\
   0 & {\mbox{\small{$\frac12$}}}
   \bar{\cal D}^2 \,\bar {{\cal J}}^{\bar a}{}_{\bar b}
   \end{array}\right),
\label{jacobian}
\end{equation}
with its antichiral and chiral sectors given by
$$
  {\cal J}_a{}^b= \left(\begin{array}{cc}
  \epsilon^{\underline{\alpha}}
  \nabla_{\underline{\alpha}} \,
  \delta_\alpha{}^\beta & \epsilon_\alpha \\
  0& \epsilon^{\underline{\alpha}}
  \nabla_{\underline{\alpha}}
  \end{array}\right), \quad\quad
  \bar {{\cal J}}^{\bar a}{}_{\bar b}=
  \left(\begin{array}{cc}
  \epsilon^{\underline{\alpha}}
  \nabla_{\underline{\alpha}} \,
  \delta^{\dot{\alpha}}{}_{\dot{\beta}}
  & \bar\epsilon^{\dot{\alpha}} \\
  0& \epsilon^{\underline{\alpha}}
  \nabla_{\underline{\alpha}}
\end{array}\right).
$$

In an analogous way, the hessian of the interaction term in
(\ref{compaction}) results in
$I_A{}^B= (P_q {\cal I} P_q)_A{}^B$,
with the `naive' hessian ${\cal I}_A{}^B$ expressed as
$$
  {\cal I}_A{}^B=
  \left(\begin{array}{cccc}
   b_1 \, \delta_\alpha{}^\beta \bar K
   &  b_1\, \theta_\alpha \bar K &
   0 & b_1\, G_\alpha\\
   b_1\, \theta^\beta \bar K&b_1\,
   \theta^2 \bar K&{\bar b_1}\,\bar G_{\dot{\beta}}&
\left(b_1\, G^\alpha\theta_\alpha
+  {\bar b_1}\, \bar G_{\dot{\alpha}}
\bar\theta^{\dot{\alpha}}\right)\\
0& {\bar b_1}\, \bar G^{\dot{\alpha}} &
\bar b_1 \, \delta^{\dot{\alpha}}{}_{\dot{\beta}} K
&  \bar b_1\, \bar \theta^{\dot{\alpha}} K \\
   b_1 \, G^\beta &
\left(b_1\, G^\alpha\theta_\alpha+  {\bar b_1}\,
\bar G_{\dot{\alpha}}\bar\theta^{\dot{\alpha}}\right)&
   \bar b_1\, \bar \theta_{\dot{\beta}} K &
   \bar b_1\, \bar \theta^2 K
  \end{array}\right).
$$
Finally, the kinetic operator is found
to be $F_A{}^B= (P_q {\cal F} P_q)_A{}^B$,
with the `naive' kinetic term ${\cal F}_A{}^B$
given by
$$
   {\cal F}_A{}^B=\left(\begin{array}{cccc}
   0&0& i a_1\partial_{\alpha{\dot{\beta}}}
   & i a_1\partial_{\alpha{\dot{\beta}}}
   \bar\theta^{\dot{\beta}}\\
   0&0& i a_1\theta^\alpha
   \partial_{\alpha{\dot{\beta}}}&a_2+
   ia_1\theta\partial\bar\theta\\
   i a_1\partial^{{\dot{\alpha}}\beta}
   & i a_1\partial^{{\dot{\alpha}}\beta}
   \theta_\beta&0&0\\
   i a_1\bar\theta_{\dot{\alpha}}
   \partial^{{\dot{\alpha}}\beta}
   &a_2+ia_1\bar\theta\partial\theta&0&0
   \end{array}\right).
$$

Introducing then as operator $T^{-1}$
the free propagator of the model in
superspace up to $(-\Box)^{-1}$, namely
$(T^{-1})_A{}^B=(P_q {\cal T}^{-1} P_q)_A{}^B$ with
$$
  ({\cal T}^{-1})_A{}^B= \frac{-1}{4(a_1+a_2)}
  \left(\begin{array}{cccc}
   0&0& \left(\frac{i (a_1+2 a_2)}{2a_1\Box}
   \partial_{\alpha{\dot{\beta}}}
   -\theta_\alpha\bar\theta_{\dot{\beta}} \right)
   & \theta_\alpha\\
   0 & 0 & \bar\theta_{\dot{\beta}}& -1 \\
   \left(\frac{i (a_1+2 a_2)}{2a_1\Box}
   \partial^{{\dot{\alpha}}\beta}
   -\bar\theta^{\dot{\alpha}}\theta^\beta \right)
   & \bar\theta^{\dot{\alpha}}&0&0\\
   \theta^\beta& -1& 0&0
   \end{array}\right),
$$
a suitable regulator, diagonal and quadratic in space-time
derivatives, arises
$$
   R_A{}^B= -\Box \, (P_q)_A{}^B.
$$
In this way the corresponding smearing and shadow
kinetic
operators (\ref{smearing op}), (\ref{shadow kinetic op}),
adapted to the
chiral case, result in
$$
  ({\cal{\varepsilon}}^2)_A{}^B= {\cal{\varepsilon}}^2 \,
  (P_q)_A{}^B,\quad\quad
   {\cal O}_A{}^B= \hat\sigma\,
   (P_q {\cal T}^{-1} P_q)_A{}^B\ ,
$$
with ${\cal{\varepsilon}}^2$ and $\hat\sigma$ defined as
$$
  {\cal{\varepsilon}}^2= \exp{(-\Box/\Lambda^2)},
  \quad\quad
  \hat\sigma=\int_0^1 \frac {dt}{\Lambda^2}
  \exp{(-t\Box/\Lambda^2)}.
$$

The form of the candidate anomalies
(\ref{can5}) --involving only
either products of
antichiral fields $H^\alpha$, $K$, or of
chiral fields
$\bar H^{\dot{\alpha}}$, $\bar K$, but no
crossed terms--
indicates that the evaluation of their
coefficients by means of the
supertrace (\ref{gen anom 2}) can now be
considerably simplified by
considering for instance only the antichiral
sector, i.e.\, by
neglecting the fields
$\bar H^{\dot{\alpha}}$ and $\bar K$, and by
further restricting the computation to
only linear and trilinear terms in $H^\alpha$,
$K$, namely to the first
and third order interaction terms%
\footnote{This
restriction is indeed sufficient
even though candidate
anomalies are defined only modulo
trivial solutions of the
consistency conditions. The reason
is that the supersymmetry
transformations of table 1 are linear
and do not mix the fields
of the chiral and antichiral sector.}.
The coefficients coming from the chiral
sector contributions can then be
automatically determined by complex conjugation.
Therefore, from now on we
are going to concentrate our attention in the terms
\begin{equation}
  \tilde {\cal A}_n=
  \left[(-1)^{A} ({{\cal{\varepsilon}}}^2)_A{}^B\,
  J_B{}^C \,\left({{\cal O}}_C{}^D I_D{}^A\right)^{n}
  \right]_{\rm anti},
  \quad\quad \mbox{for $n=1,3$},
\label{relevant trace}
\end{equation}
where the subscript `anti' indicates that all
terms involving
$\bar H^{\dot{\alpha}}$ and $\bar K$ are neglected.

Our main task shall now consist in determining
the diagonal entries of
the matrix involved in expression
(\ref{relevant trace}). First of all,
the $n$th power of the matrix
${\cal O}_A{}^C I_C{}^B$ reads,
under the above restrictions
$$
  ({\cal O}_A{}^C I_C{}^B)_{\rm anti}^n=
   \left(\begin{array}{cc}
   ({\cal O} I_n)_a{}^b& \cdots \\
   0  & ( {{\cal O} I}_n)^{\bar a}{}_{\bar b}
   \end{array}\right).
$$
Its diagonal blocks --the relevant
ones taking into account the block
diagonal form of the jacobian
(\ref{jacobian})-- can be easily found
by using the commutation relation
\begin{equation}
  \left[{\mbox{\small{$\frac12$}}}{\cal D}^2,
  \theta_\alpha\right]= {\cal D}_\alpha,
\label{commutators}
\end{equation}
resulting in
\begin{eqnarray}
  ({\cal O} I_n)_a{}^b&=& {\mbox{\small{$\frac12$}}}
  {\cal D}^2\left(\begin{array}{cc}
       \theta_\alpha (S^\gamma{\cal D}_\gamma)^{n-1} &
       -\theta_\alpha (S^\gamma{\cal D}_\gamma)^{n-1}
       \theta_\beta\\
       -(S^\gamma{\cal D}_\gamma)^{n-1}  &
       (S^\gamma{\cal D}_\gamma)^{n-1}\theta_\beta
   \end{array}\right)S^\beta {\mbox{\small{$\frac12$}}}
   {\cal D}^2,
\nonumber\\
  ({{\cal O} I}_n)^{\bar a}{}_{\bar b}&=&
            {\mbox{\small{$\frac12$}}}
            \bar{\cal D}^2\left(\begin{array}{cc}
             0 & \cdots \\
             0 & ({\cal D}_\gamma {\cal G}^\gamma)^n
   \end{array}\right){\mbox{\small{$\frac12$}}}
   \bar{\cal D}^2,
\nonumber
\end{eqnarray}
in terms of the quantities $S^\alpha$,
${\cal G}^\alpha$ defined as
\begin{equation}
    S^\alpha={\mbox{\small{$\frac12$}}}
    \bar{\cal D}^2 {\cal G}^\alpha,
   \quad\quad
   {\cal G}^\alpha=
   \left(\frac{-b_1}{4(a_1+a_2)}\right)
   \hat\sigma G^\alpha,
\label{sdef}
\end{equation}
where all the operators are understood
to act on everything on
their right. Terms indicated by dots
in the above matrices
turn out to be irrelevant for the present computation.

Afterwards, straightforward matrix multiplication yields
$$
  {\rm diag}\,\left(({\cal{\varepsilon}}^2)_A{}^B\,
  J_B{}^C \,\left({\cal O}_C{}^D
  I_D{}^E\right)^{n}\right)_{\rm anti}=
  \left( (A_n)_\alpha{}^\beta, A_n; 0, C_n \right),
$$
where the expressions for the
antichiral sector operators are
found to be, upon use of the
commutation relation
$\left[\epsilon^{\underline{\beta}}
\nabla_{\underline{\beta}},
\theta_\alpha\right]= \epsilon_\alpha$,
\begin{eqnarray}
     (A_n)_\alpha{}^\beta&=&
     {\cal{\varepsilon}}^2 \,
     {\mbox{\small{$\frac12$}}} {\cal D}^2
     \left[\epsilon^{\underline{\delta}}
     \nabla_{\underline{\delta}}\,
     \theta_\alpha - \epsilon_\alpha\right]
     (S^\gamma {\cal D}_\gamma)^{n-1} \,
     S^\beta\, {\mbox{\small{$\frac12$}}} {\cal D}^2
\nonumber\\
     &=&{\cal{\varepsilon}}^2 \,
     {\mbox{\small{$\frac12$}}} {\cal D}^2\,
     \theta_\alpha \,
     \epsilon^{\underline{\delta}}
     \nabla_{\underline{\delta}}\,
     (S^\gamma {\cal D}_\gamma)^{n-1} \,
     S^\beta\, {\mbox{\small{$\frac12$}}} {\cal D}^2,
\nonumber\\
     A_n&=& {\cal{\varepsilon}}^2\,
     {\mbox{\small{$\frac12$}}} {\cal D}^2\,
     \epsilon^{\underline{\delta}}
     \nabla_{\underline{\delta}}\,
     (S^\gamma {\cal D}_\gamma)^{n-1}\,
     \theta_\beta\, S^\beta\,
     {\mbox{\small{$\frac12$}}}{\cal D}^2,
\label{acs}
\end{eqnarray}
whereas the chiral sector operator is directly given by
\begin{equation}
   C_n= {\cal{\varepsilon}}^2 \,
   \epsilon^{\underline{\delta}}
   \nabla_{\underline{\delta}}\,
  {\mbox{\small{$\frac12$}}}
  \bar{\cal D}^2\, {\cal D}_{\alpha_1}\,
  {\cal G}^{\alpha_1} {\mbox{\small{$\frac12$}}}
  \bar{\cal D}^2
  \ldots
  {\mbox{\small{$\frac12$}}} \bar{\cal D}^2 \,
  {\cal D}_{\alpha_n}\,
  {\cal G}^{\alpha_n}{\mbox{\small{$\frac12$}}}
  \bar{\cal D}^2.
\label{csc}
\end{equation}
The general expression of $\tilde {\cal A}_n$
(\ref{relevant trace}) is thus
\begin{equation}
  \tilde {\cal A}_n=
  {\rm Tr}\left[ - (A_n)_\alpha{}^\alpha + A_n\right]
               +{\rm \overline Tr} [C_n],
\label{anom funct}
\end{equation}
where the extra minus sign comes from
taking the discrete trace over the
fermionic fields, while the symbols Tr
and ${\rm \overline Tr}$ stand
respectively for the functional traces
in the antichiral and chiral
superspaces, namely
\begin{equation}
    {\rm Tr} [A] =\int{\rm d}^6 \bar z\,
    \left.{A(z, z')}\right|_{\bar z=\bar z'},
    \quad\quad
    {\rm \overline Tr} [C] =\int{\rm d}^6 z\,
    \left.{C(z, z')}\right|_{z=z'}.
\label{supertraces}
\end{equation}

Upon substitution of expressions
(\ref{acs}) and (\ref{csc}), both
traces in (\ref{anom funct}) are then
seen to share similar structures.
However, there is the fundamental
difference that such functional traces
are taken in different superspaces,
according to (\ref{supertraces}).
Therefore, in order to compare both
expressions, some mechanism should be
found to relate supertraces of antichiral
expressions to those of chiral
ones. Fortunately, it is not difficult to
verify, as shown in appendix
\ref{aptrick}, that for chiral operators
$\bar{\cal A}$, namely those
verifying $\bar {\cal D}_{\dot{\alpha}} \bar {\cal A}= 0$,
the following relation holds
\begin{equation}
    {\rm Tr}\left[ {\mbox{\small{$\frac12$}}}
    {\cal D}^2 \,\bar{\cal A}\,
    {\mbox{\small{$\frac12$}}} {\cal D}^2\right]=
    {\rm \overline Tr}\left[\bar{\cal A}\,
    {\mbox{\small{$\frac12$}}} {\cal D}^2 \,
    {\mbox{\small{$\frac12$}}}\bar{\cal D}^2\right].
\label{trick}
\end{equation}
Using this result as well as the commutation relation
(\ref{commutators}) and the cyclic property
of the regulated trace, the
antichiral sector contribution
${\rm Tr}\left[ - (A_n)_\alpha{}^\alpha + A_n\right]$
to (\ref{anom funct}) can be
rewritten in chiral form as
\begin{equation}
  {\rm Tr}\left[ - (A_n)_\alpha{}^\alpha + A_n\right]=
  {\rm \overline Tr} [B_n]\ ,
\label{fasc}
\end{equation}
with the operator $B_n$ given by
$$
    B_n= {\cal{\varepsilon}}^2 \,
    \epsilon^{\underline{\delta}}
    \nabla_{\underline{\delta}}\,
     {\mbox{\small{$\frac12$}}}
     \bar{\cal D}^2 \,{\cal G}^{\alpha_1}\,
     {\cal D}_{\alpha_1} {\mbox{\small{$\frac12$}}}
     \bar {\cal D}^2
     \ldots
     {\mbox{\small{$\frac12$}}} \bar{\cal D}^2 \,
     {\cal G}^{\alpha_n}\, {\cal D}_{\alpha_n}
     {\mbox{\small{$\frac12$}}} \bar {\cal D}^2,
$$
after substitution of $S^\alpha$ by
its explicit expression (\ref{sdef}). In
this way, $B_n$ is seen to `almost'
coincide with $C_n$ (\ref{csc}) when
reading it from the right to the left.

This similarity may conveniently be
exploted by using
the property that the trace of an
operator and of its transpose coincide.
Combining further this fact with
the cyclic property of regulated
traces, the following relations are seen to hold
\begin{eqnarray}
  {\rm \overline Tr} [B_n]&=&
  {\rm \overline Tr} \left[
  {\cal{\varepsilon}}^2 \,
  \epsilon^{\underline{\delta}}
  \nabla_{\underline{\delta}}\,
  {\mbox{\small{$\frac12$}}}
  \bar{\cal D}^2 \,{\cal G}^{\alpha_1}\,
  {\cal D}_{\alpha_1} {\mbox{\small{$\frac12$}}}
  \bar {\cal D}^2
  \ldots
  {\mbox{\small{$\frac12$}}} \bar{\cal D}^2
  \,{\cal G}^{\alpha_n}\, {\cal D}_{\alpha_n}
  {\mbox{\small{$\frac12$}}}\bar{\cal D}^2\right]
\nonumber\\
  &=& {\rm \overline Tr} \left[
  {\mbox{\small{$\frac12$}}} \bar{\cal D}^2
  \,{\cal G}^{\alpha_1}\, {\cal D}_{\alpha_1}
  {\mbox{\small{$\frac12$}}} \bar {\cal D}^2
  \ldots
  {\mbox{\small{$\frac12$}}} \bar{\cal D}^2
  \,{\cal G}^{\alpha_n}\, {\cal D}_{\alpha_n}
  {\mbox{\small{$\frac12$}}}\bar{\cal D}^2\,
  \epsilon^{\underline{\delta}}
  \nabla_{\underline{\delta}}\,
  {\cal{\varepsilon}}^2 \right]
\nonumber\\
  &=& -(-)^{2n} {\rm \overline Tr} \left[
  {\cal{\varepsilon}}^2 \,
  \epsilon^{\underline{\delta}}
  \nabla_{\underline{\delta}}\,
  {\mbox{\small{$\frac12$}}}
  \bar{\cal D}^2\, {\cal D}_{\alpha_1}\,
  {\cal G}^{\alpha_1} {\mbox{\small{$\frac12$}}}
  \bar{\cal D}^2
  \ldots
{\mbox{\small{$\frac12$}}} \bar{\cal D}^2 \,
{\cal D}_{\alpha_n}\,
{\cal G}^{\alpha_n}
{\mbox{\small{$\frac12$}}}\bar{\cal D}^2\right]
  = - {\rm \overline Tr} [C_n]\ ,
\nonumber
\end{eqnarray}
so that the contribution coming from
the antichiral sector,
${\rm \overline Tr} [B_n]$ (\ref{fasc}),
is seen to exactly
cancel that coming from the chiral sector,
${\rm \overline Tr} [C_n]$, for all
$n$. The present computation leads
thus to the vanishing of
$\tilde {\cal A}_n$ (\ref{anom funct})
for all $n$ and with it, of
the potential anomalies of our model.
Therefore, we conclude that the latter,
potentially present on
cohomological grounds, actually do not show up in
the model we have analyzed at the one loop level.
We have also checked that this remains valid for
supersymmetric actions which differ from (\ref{simpleact}) 
and arise from (\ref{a1}) 
by turning on other (combinations of) coefficients
such as $a_3$, $m$ or $b_3$. However, we have not performed
the computation for the most general action (\ref{a1}), as
the main purpose of considering the toy model was the
illustration of the method outlined in section \ref{reg}.

\section{Conclusion}
\label{conclusion}

\hspace{\parindent}%
The purpose of this paper is to show
that implementation of superspace
techniques in the framework of nonlocal
regularization constitutes a
suitable and efficient tool to analyze
anomaly issues.
To outline and illustrate the method,
we have applied it
to a toy model whose supersymmetry,
by cohomological arguments, is
potentially anomalous, but turns out
to be actually nonanomalous at the
one-loop level. As a byproduct, the
result of the computation
gives further evidence that the
remarkable quantum stability
of supersymmetry even extends
to models which admit nontrivial
solutions of the consistency condition
for supersymmetry anomalies.
Finally, although not proven, our construction also
points out to nonlocal regularization
as a possible candidate for a
supersymmetric invariant regularization method.

\section*{Acknowledgement.}
This work was carried
out in the framework of the European
Community Research Programme
``Gauge theories, applied supersymmetry
and quantum gravity", with a
financial contribution under contract
SC1-CT92-0789.

\appendix

\section{Conventions and notation}\label{conv}

\subsection{Lorentz ($SL(2,{\bf C})$) invariant tensors}
\label{app1}

Minkowski metric, $\varepsilon$-tensors:
\begin{eqnarray*} & &\eta_{ab}=diag(1,-1,-1,-1),\quad
\varepsilon^{abcd}=\varepsilon^{[abcd]},
\quad \varepsilon^{0123}=1,\\
& &\varepsilon^{\alpha\beta}=
-\varepsilon^{\beta\alpha},\quad
\varepsilon^{{\dot{\alpha}}{\dot{\beta}}}=
-\varepsilon^{{\dot{\beta}}{\dot{\alpha}}},\quad
\varepsilon^{12}=\varepsilon^{\dot 1\dot 2}=1,\\
& &\varepsilon_{\alpha\gamma}
\varepsilon^{\gamma\beta}=
\delta_\alpha{}^\beta=diag(1,1),\quad
\varepsilon_{{\dot{\alpha}}\dot{\gamma}}
\varepsilon^{\dot{\gamma}{\dot{\beta}}}=
\delta^{\dot{\beta}}{}_{\dot{\alpha}}=diag(1,1)
\end{eqnarray*}
$\sigma$--matrices:
$\sigma^a{}_{\alpha{\dot{\beta}}}$
($\alpha$: row index,
${\dot{\beta}}$: column index):
\[
\sigma^0=\left( \begin{array}{rr}1&0\\
0&1\end{array}\right),\quad
\sigma^1=\left( \begin{array}{rr}0&1\\
1&0\end{array}\right),\quad
\sigma^2=\left( \begin{array}{rr}0&-i\\
i&0\end{array}\right),\quad
\sigma^3=\left( \begin{array}{rr}1&0\\
0&-1\end{array}\right)  \]
$\bar \sigma$--matrices:
\begin{eqnarray*}
\bar \sigma^{a\, {\dot{\alpha}}\alpha}=
\varepsilon^{{\dot{\alpha}}{\dot{\beta}}}
\varepsilon^{\alpha\beta}\sigma^a{}_{\beta{\dot{\beta}}}
\end{eqnarray*}
$\sigma^{ab},\bar \sigma^{ab}$--matrices:
\begin{eqnarray*} \sigma^{ab}{}_\alpha{}^\beta =
{\mbox{\small{$\frac14$}}}(\sigma^a\bar \sigma^b
-\sigma^b\bar \sigma^a)_\alpha{}^\beta,\quad
\bar \sigma^{ab}{}^{\dot{\alpha}}{}_{\dot{\beta}}
={\mbox{\small{$\frac14$}}}(\bar{\sigma}^a\sigma^b-
\bar \sigma^b\sigma^a)^{\dot{\alpha}}{}_{\dot{\beta}}
\end{eqnarray*}

\subsection{Spinors, grading and complex conjugation}
\label{app2}

We work with two-component Weyl spinors.
Undotted and dotted
spinor indices $\alpha,{\dot{\alpha}}$
distinguish
the $({\mbox{\small{$\frac12$}}},0)$ and
$(0,{\mbox{\small{$\frac12$}}})$ representations
of $SL(2,{\bf C})$ related by complex conjugation.

\noindent
Raising and lowering of spinor indices:
\begin{eqnarray*} \psi_\alpha=
\varepsilon_{\alpha\beta}\psi^\beta,\quad
\psi^\alpha=\varepsilon^{\alpha\beta}\psi_\beta,\quad
\bar \psi_{\dot{\alpha}}=
\varepsilon_{{\dot{\alpha}}{\dot{\beta}}}
\bar \psi^{\dot{\beta}},\quad \bar \psi^{\dot{\alpha}}
=\varepsilon^{{\dot{\alpha}}{\dot{\beta}}}
\bar \psi_{\dot{\beta}}
\end{eqnarray*}
Contraction of spinor indices:
\begin{eqnarray*} \psi\chi:=
\psi^\alpha\chi_\alpha,\quad \bar \psi\bar \chi:
=\bar \psi_{\dot{\alpha}}\bar \chi^{\dot{\alpha}}
\end{eqnarray*}
Lorentz vector indices in spinor notation:
\begin{eqnarray*} V_{\alpha{\dot{\alpha}}}=
\sigma^a{}_{\alpha{\dot{\alpha}}} V_a
\end{eqnarray*}

The grading (Grassmann parity) $|X|$
of a field or an operator $X$
is determined by the number of its
spinor indices and its ghost
number ($gh$),
\[ |X_{\alpha_1\ldots\alpha_n}^{{\dot{\alpha}}_1
\ldots{\dot{\alpha}}_m} |=
m+n+gh(X)\quad (mod\ 2).                        \]
The grading of the fields $\phi^i$ determines
their statistics,
\begin{eqnarray*}
\phi^i\phi^j=(-)^{|\phi^i|\, |\phi^j|}\phi^j\phi^i.
\end{eqnarray*}
Complex conjugation of a field or operator $X$
is denoted by $\bar X$.
Complex conjugation of products of fields
and operators is defined by
\begin{eqnarray*}
\overline{XY}=(-)^{| X|\, | Y|}\bar X\, \bar Y.
\end{eqnarray*}
In particular this implies
\begin{eqnarray*}\overline{\partial /
\partial \phi}=(-)^{|\phi|}\partial /
\partial \bar \phi
\end{eqnarray*}
and thus the minus sign in front of
$\partial /\partial \bar \theta$ in
(\ref{sf1}) and (\ref{sf3}).

\subsection{Superspace conventions and useful identities}
\label{app3}

$\theta^\alpha$ and $\bar \theta^{\dot{\alpha}}$
are odd graded, constant and
related by complex conjugation.

\noindent
Superspace integration:
\begin{eqnarray*}
& & \int d\theta\, \theta=\int d\bar \theta\,
\bar \theta=1,\quad
\int d^2\theta=\int d\theta^2d\theta^1 ,\\
& &\int d^2\bar \theta=
\int d\bar \theta^{\dot 1}d\bar \theta^{\dot 2}
,\quad
\int d^4\theta=\int d^2\theta d^2\bar \theta,\\
& &\int d^6z=\int d^4x d^2\theta,\quad
\int d^6\bar z=\int d^4x d^2\bar \theta,\quad
\int d^8z=\int d^4x d^4\theta
\end{eqnarray*}
$\delta$-functions:
\begin{eqnarray*}
\delta^2(\theta-\theta')&=&
-{\mbox{\small{$\frac12$}}}(\theta-\theta')^2\\
\delta^2(\bar \theta-\bar \theta')&=&
-{\mbox{\small{$\frac12$}}}(\bar \theta-\bar \theta')^2\\
\delta^6(z-z')&=&\delta^2(\theta-\theta')\delta^4(x-x')\\
\delta^6(\bar z-\bar z')&=&
\delta^2(\bar \theta-\bar \theta')\delta^4(x-x')\\
\delta^8(z-z')&=&
\delta^2(\theta-\theta')
\delta^2(\bar \theta-\bar \theta')\delta^4(x-x')
\end{eqnarray*}
Useful identities:
\begin{eqnarray}
\exp(\theta D+\bar \theta\bar D)&=&
\exp(i\theta\partial \bar \theta)
\exp(\theta D)\exp(\bar \theta\bar D)\nonumber\\
&=&\exp(-i\theta\partial \bar \theta)
\exp(\bar \theta\bar D)\exp(\theta D)
\label{useful}\\
{{\cal D}}_{\underline{\alpha}}
\exp(\theta D+\bar \theta \bar D)&=&
\exp(\theta D+\bar \theta \bar D)\,
D_{\underline{\alpha}}.
\nonumber
\end{eqnarray}
$\theta$-integrations over superfields
(\ref{Bdef}) result thus in
\begin{eqnarray}
& &\int d^2\theta
\exp(\theta D+\bar \theta \bar D)
f(\phi,\partial \phi,\ldots)\cong
{\mbox{\small{$\frac12$}}}D^2
\exp(\bar \theta \bar D)\, f(\phi,\partial \phi,\ldots),
\nonumber\\
& & \int d^4\theta
\exp(\theta D+\bar \theta \bar D)
f(\phi,\partial \phi,\ldots)\cong
{\mbox{\small{$\frac14$}}}D^2
\bar D^2f(\phi,\partial \phi,\ldots)
\nonumber
\end{eqnarray}
where $\cong$ denotes equality
up to a total derivative.

\section{Superfields and constituents}
\label{disc}

\hspace{\parindent}
In this appendix, we briefly review
the construction of
superfields out of ordinary fields
for given supersymmetry
transformations of the latter
according to the conventions used
in this paper.
As usual we implement the
supersymmetry transformations
on superfields  through the operators
$\nabla_\alpha$,
$\bar \nabla_{\dot{\alpha}}$ (\ref{sf1}).
Then, given a (linear) representation
$D_\alpha,\bar D_{\dot{\alpha}}$
 of the supersymmetry
algebra (\ref{susyalg})  on ordinary fields $\phi^i$ such
as in table 1 of section \ref{mult}, superfields are defined as functions
$\Sigma$
of the $\theta^\alpha$,
$\bar \theta^{\dot{\alpha}}$, $\phi^i$
and of the derivatives of the $\phi^i$,
$\Sigma=\Sigma(\theta,
\bar \theta,\phi,\partial \phi,\ldots)$,
satisfying
\begin{equation}
    D_\alpha \Sigma=\nabla_\alpha \Sigma\ ,\quad
    \bar D_{\dot{\alpha}}
    \Sigma=\bar \nabla_{\dot{\alpha}} \Sigma\ ,
\label{sf2}
\end{equation}
where $D_\alpha$ and $\bar D_{\dot{\alpha}}$ act
nontrivially only on the $\phi^i$
and their derivatives
and anticommute with all the
$\theta$'s and $\bar \theta$'s.
The operators $\nabla_\alpha$,
$\bar \nabla_{\dot{\alpha}}$ (\ref{sf1})
provide then a representation of the
supersymmetry algebra (\ref{susyalg}) with
$(P_a,Q_\alpha,\bar Q_{\dot{\alpha}})
\equiv (-\partial_a,-\nabla_\alpha,
-\bar \nabla_{\dot{\alpha}})$.
Note that $\nabla_\alpha \Sigma$ is {\em not}
a superfield since its
$\bar D_{\dot{\alpha}}$ transformation is not
given by
$\bar \nabla_{\dot{\alpha}}\nabla_\alpha\Sigma$,
but rather by
\[ \bar D_{\dot{\alpha}}\nabla_\alpha\Sigma=
-\nabla_\alpha\bar D_{\dot{\alpha}}\Sigma
=-\nabla_\alpha\bar \nabla_{\dot{\alpha}}\Sigma\ .\]
Instead, and in contrast to the $\nabla$'s,
the standard `covariant
derivatives ${{\cal D}}_\alpha$,
$\bar {\cal D}_{\dot{\alpha}}$ (\ref{sf3})
map superfields to superfields because
they anticommute both
with the $D$'s and with the $\nabla$'s.

Having characterized superfields
abstractly by (\ref{sf2}), we can now
construct them explicitly:
any superfield, i.e.\ any solution of
(\ref{sf2})
can be written in the form
\begin{equation}
\Sigma=\exp(\theta D+\bar \theta \bar D)\,
f(\phi,\partial \phi,\ldots)\ ,
\label{Bdef}
\end{equation}
where $f(\phi,\partial \phi,\ldots)$ is a
function of the (ordinary)
fields and their derivatives and we used
the summation conventions
$\theta D=\theta^\alpha D_\alpha$ and
$\bar \theta\bar D=\bar \theta_{\dot{\alpha}}
\bar D^{\dot{\alpha}}$.
The proof of this statement is straightforward using that
(i) (\ref{Bdef}) satisfies (\ref{sf2}) for
any $f(\phi,\partial \phi,\ldots)$, as can be
easily checked directly, (ii)
any nonvanishing superfield has a nonvanishing
$\theta$-independent part which is required by
(\ref{sf2}).
The assertion is now proved as follows:
given a nonvanishing solution $\Sigma$ of
(\ref{sf2}) with $\theta$-independent part
$f(\phi,\partial \phi,\ldots)$ we
consider $\Sigma'=\Sigma-\exp(\theta D
+\bar \theta \bar D)f(\phi,\partial \phi,\ldots)$.
The latter is a superfield due to (i)
and must vanish due to (ii)
since by construction it has no
$\theta$-independent part.

\section{Lagrangian and candidate anomaly in explicit form}
\label{appact}

The various parts (\ref{a2}--\ref{a5}) of the general
Lagrangian read
explicitly (up to total derivatives)
\begin{eqnarray*}
\int d^2\bar \theta\, K&\cong&-F,\\
\int d^4\theta\,  iG\partial \bar  G
&\cong&
i\eta\partial \bar \eta
-i\psi\partial \bar \psi+
2\psi\Box\chi+2\bar \psi\Box\bar \chi-
4i\chi\Box\partial \bar \chi-4A\Box\bar A\\
&& -4(\partial_aV^a)\partial_b\bar V^b
+2F_{ab}\bar F^{ab}
+2iF\partial_a\bar V^a-2i\bar F\partial_a V^a,\\
\int d^4\theta\,  K\bar  K
&\cong&-4A\Box \bar A-2i
\bar \psi\partial \psi+F\bar F,\\
\int d^4\theta\,
{\mbox{\small{$\frac14$}}}G\bar {\cal D}^2 G
&\cong&
2i\eta\partial \bar \psi+
4\eta\Box\chi-4V_a\Box V^a
-F^2-4iF\partial_a V^a,\\
\int d^4\theta\,  GG
&\cong&
-2\bar \psi\bar \psi+4F A,\\
\int d^4\theta\,
{\mbox{\small{$\frac12$}}}G G\bar  K &\cong&
-\bar A\bar \psi\bar \psi
-2 \chi\chi\Box\bar A+
2i\chi\sigma^a\bar \psi\partial_a\bar A
-A\eta\psi\\
& &+V^{\alpha{\dot{\alpha}}}
(\bar \psi_{\dot{\alpha}}
\psi_\alpha
-2i\chi_\alpha\partial_{\beta{\dot{\alpha}}}\psi^\beta)
-4iAV^a\partial_a\bar A\\
& &-\bar F(V^aV_a-\chi\eta)
+F(2A\bar A-\chi\psi),\\
\int d^4\theta\, {\mbox{\small{$\frac16$}}}
G GK &\cong&
-A\bar \psi\bar \psi+A^2F,\\
\int d^4\theta\,  {\mbox{\small{$\frac14$}}}G G\,
\bar G\bar G &\cong&
     {\mbox{\small{$\frac12$}}}
     A^2\bar A^2+V^a\bar V_aA\bar A+
     {\mbox{\small{$\frac12$}}}V^aV_a\bar V^b\bar V_b
     -F(\bar \chi \bar V\chi-A\bar \chi\bar \chi)
\\ & &
     -\chi\eta\bar V_a\bar V^a+\chi V \bar \eta \bar A
     -\chi\sigma^a\bar \sigma^b\psi V_a \bar V_b
     -\chi \bar V \bar \psi \bar A - 2\chi\psi A\bar A
\\ & &
     +i(A\chi)\partial  (\bar \chi \bar A)-
     2iAV^a\partial_a(\bar \chi\bar \chi)
     -i(V^{{\dot{\alpha}}\beta}\chi_\beta)
     \partial_{\alpha{\dot{\alpha}}}
     (\bar \chi_{\dot{\beta}} \bar V^{{\dot{\beta}}\alpha})
\\ & &
     +{\mbox{\small{$\frac12$}}}\chi\eta\, \bar \chi\bar \eta
     +{\mbox{\small{$\frac12$}}}\chi\psi\, \bar \chi\bar \psi
     -{\mbox{\small{$\frac12$}}}(\chi\chi)(\psi\psi)
\\ & &
      +i\chi\chi\partial_a(\psi\sigma^a\bar \chi)
      -(\chi\chi)\Box(\bar \chi\bar \chi)+c.c.
\end{eqnarray*}
with $F_{ab}=\partial_aV_b-\partial_bV_a$ and
$\Box=\partial_a\partial^a=
{\mbox{\small{$\frac12$}}}
\partial_{\alpha{\dot{\alpha}}}
\partial^{{\dot{\alpha}}\alpha}$.

The integrand of the candidate
anomaly $\Delta_2$ in (\ref{can5}) reads explicitly
\begin{eqnarray*}
{\mbox{\small{$\frac12$}}}
\bar D^2(\xi\chi\bar \psi'\bar \psi')=
\xi\chi\{2i \varepsilon^{abcd}F_{ab}F_{cd}+
8(\partial_aV^a)\partial_b V^b+
4F_{ab}F^{ab}-2F^2\\
-8iF\partial_a V^a
+4i\bar \psi'\partial \eta\}
-\xi\eta\bar \psi'\bar \psi'
-2\xi V\bar \psi'F-
4i\xi \sigma^a\bar \sigma^b\sigma^c
\bar \psi' V_a\partial_cV_b
\end{eqnarray*}

\section{Proof of relation (4.11)}
\label{aptrick}

\hspace{\parindent}%
In the perturbative computation of
the anomaly coefficients performed in
section \ref{comp}, relation
(\ref{trick}) has been seen to be crucial in
checking their vanishing. In this appendix,
we prove that relation.

Consider a generic chiral operator $\bar{\cal A}$,
namely an object
verifying $\bar {\cal D}_{\dot{\alpha}} \bar {\cal A}= 0$,
and a typical trace over this
quantity of the form
\begin{eqnarray}
  {\rm Tr}\left[ {\mbox{\small{$\frac12$}}} {\cal D}^2 \,
  \bar{\cal A}\, {\mbox{\small{$\frac12$}}}
  {\cal D}^2\right]&=&
  \int{\rm d}^6 \bar z\,
    \left.{\left[ {\mbox{\small{$\frac12$}}}
    {\cal D}^2 \,\bar{\cal A}\,{\mbox{\small{$\frac12$}}}
    {\cal D}^2
    \delta^8(z-z') \right]} \right|_{\bar z=\bar z'}
\nonumber\\
  &=&\int{\rm d}^6 \bar z\,{\rm d}^6 \bar z'
  \left[{\mbox{\small{$\frac12$}}}{\cal D}^2 \,
  \bar{\cal A}\,{\mbox{\small{$\frac12$}}}
  {\cal D}^2 \delta^8(z-z')\right]
  \left[{\mbox{\small{$\frac12$}}}{\cal D}^2 \,
  \delta^8(z'-z)\right].
\label{tr1}
\end{eqnarray}
The identity for chiral expressions
$$
  \bar{\cal A}(z)= \int{\rm d}^6 z''\,
  {\mbox{\small{$\frac12$}}} \bar{\cal D}^2\,
  \delta^8(z-z'')\,\bar{\cal A}(z''),
$$
allows to rewrite (\ref{tr1}) as
\begin{eqnarray}
  &&\int{\rm d}^6 \bar z\,{\rm d}^6
  \bar z' {\rm d}^6 z''
  \left[{\mbox{\small{$\frac12$}}}
  {\cal D}^2 \,{\mbox{\small{$\frac12$}}}
  \bar{\cal D}^2 \,\delta^8(z-z'')\right]
  \left[\bar{\cal A}(z'')
  \,{\mbox{\small{$\frac12$}}}
  {\cal D}^2 \delta^8(z''-z')\right]
  \left[{\mbox{\small{$\frac12$}}}
  {\cal D}^2 \,\delta^8(z'-z)\right]=
\nonumber\\
  &&\int{\rm d}^6 \bar z {\rm d}^6 z''
  \left[\bar{\cal A}(z'')
  \,{\mbox{\small{$\frac12$}}}
  {\cal D}^2 \delta^8(z''-z)\right]
  \left[{\mbox{\small{$\frac12$}}}{\cal D}^2
  \,{\mbox{\small{$\frac12$}}}\bar{\cal D}^2
  \,\delta^8(z-z'')\right],
\label{tr2}
\end{eqnarray}
where in writing the second expression use
has been made of the property
(\ref{idempotency}) for the antichiral
projector ${\mbox{\small{$\frac12$}}}{\cal D}^2$.
By exactly the same arguments,
(\ref{tr2}) can be further rewritten as
\begin{eqnarray}
  &&\int{\rm d}^6 \bar z\,{\rm d}^6 z'
  {\rm d}^6 z''
  \left[{\mbox{\small{$\frac12$}}}
  \bar{\cal D}^2 \,\delta^8(z''-z')\right]
  \left[\bar{\cal A}(z') \,
  {\mbox{\small{$\frac12$}}}
  {\cal D}^2 \delta^8(z'-z)\right]
  \left[{\mbox{\small{$\frac12$}}}
  {\cal D}^2 \,{\mbox{\small{$\frac12$}}}
  \bar{\cal D}^2 \,\delta^8(z-z'')\right]=
\nonumber\\
  &&\int{\rm d}^6 z' {\rm d}^6 z''
  \left[{\mbox{\small{$\frac12$}}}
  \bar{\cal D}^2 \,\delta^8(z''-z')\right]
  \left[\bar{\cal A}(z'') \,
  {\mbox{\small{$\frac12$}}}
  {\cal D}^2 {\mbox{\small{$\frac12$}}}
  \bar{\cal D}^2 \,
  \delta^8(z''-z')\right]=
\nonumber\\
  &&\int{\rm d}^6 z \left.
  {\left[\bar{\cal A}(z)
  \,{\mbox{\small{$\frac12$}}}
  {\cal D}^2 {\mbox{\small{$\frac12$}}}
  \bar{\cal D}^2 \,
  \delta^8(z-z')\right]}
   \right|_{z=z'}=
  {\rm \overline Tr}
  \left[\bar{\cal A}\,{\mbox{\small{$\frac12$}}}
  {\cal D}^2 \,{\mbox{\small{$\frac12$}}}
  \bar{\cal D}^2\right],
\nonumber
\end{eqnarray}
which finally shows fulfillment of
relation (\ref{trick}).

\end{document}